\documentclass[aps,prl,twocolumn,groupedaddress,showpacs,superscriptaddress,nofootinbib]{revtex4-1}
\usepackage{graphicx}
\usepackage{color}
\usepackage{hyperref}
\begin{document}
\title{Einstein’s Cat: A Thought Experiment on the Universality of Time Dilation}

\author{V.G.~Rousseau}
\affiliation{Physics Department, Xavier University of Louisiana, 1 Drexel Dr., New Orleans, LA 70125, USA}

\relpenalty=10000      % Don't cut formulas
\binoppenalty=10000

\begin{abstract}
The universality of time dilation is sometimes misunderstood, with claims that it applies exclusively to light clocks and not to mechanical clocks or other material objects, including biological entities. This leads to incorrect interpretations of experiments like the Hafele-Keating experiment, which actually confirm relativistic effects through the use of atomic clocks. To address this issue, we propose a thought experiment, inspired by Schrödinger's cat, involving a ``Sync-or-Die clock", a device that combines a light clock and a mechanical stopwatch, with the life of Einstein's cat, Tiger, depending on their synchronization. A straightforward analysis shows that both clocks experience time dilation equally, reinforcing the fact that time dilation is not limited to clocks based on light propagation but is a universal feature of Special Relativity.
\end{abstract}

\pacs{03.30.+p, 01.40.-d, 01.55.+b}
\maketitle
\section{Introduction}
One of the most counterintuitive consequences of Special Relativity is the phenomenon of time dilation, which has been confirmed by numerous experiments. Among these, the Hafele–Keating experiment\cite{HafeleKeating} stands out as a direct and striking demonstration, showing that atomic clocks flown around the Earth experience measurable time dilation. However, despite the robust experimental support, it is not uncommon to encounter the misconception that time dilation is merely an artifact of specific clock designs, particularly those based on light propagation, such as the well-known light clock used in pedagogical derivations of the Lorentz transformation.

This misconception, that time dilation applies exclusively to clocks involving light while mechanical or other “real” clocks remain unaffected, can lead to confusion about both the interpretation of experiments and the universality of relativistic effects. Although such claims are rare in peer-reviewed literature, they do appear in some published papers challenging Special Relativity\cite{Sauerheber,Sauerheber2,Drozdzynski,Mahmud,Dixit,Ronghua}, and are widespread in informal settings such as educational forums and online discussions\cite{Forum1,Forum2,Forum3}. These arguments often rest on the incorrect assumption that derivations involving light clocks fail to generalize to other timekeeping mechanisms, thereby reinforcing the false notion that time dilation is not a genuine physical phenomenon. Addressing this misconception is therefore of clear pedagogical importance.

In this paper, we propose a simple thought experiment, inspired by Schrödinger's cat, to illustrate in a clear and pedagogical manner that time dilation applies universally, regardless of the internal mechanism of the clock. The scenario involves a hypothetical device — the Sync-or-Die clock — which compares the synchronization of a light clock and a mechanical clock, with the life of Einstein's cat, Tiger, depending on their agreement. By analyzing this setup from both an inertial frame where the device is at rest and another inertial frame where it moves at constant velocity, we show that both clocks must undergo the same time dilation, confirming that relativistic effects are independent of the type of clock used.

This thought experiment provides a concrete and accessible way to address a subtle point that often causes difficulties among students and non-experts, reinforcing the universality of relativistic time dilation.

\section{The Sync-or-Die Thought Experiment}

\begin{figure}[h]
  \centerline{\includegraphics[width=0.3\textwidth]{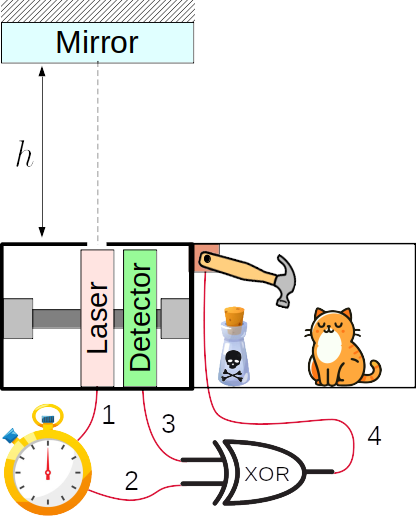}}
  \caption{The Sync-or-Die clock. The clock consists of a laser, mirror, light detector, actuator, mechanical stopwatch, XOR gate, and a poison release mechanism. When the stopwatch is started, the laser emits a light pulse toward the mirror. The actuator then shifts the laser and places the detector in position. After 30 seconds, the stopwatch signals one input of the XOR gate, and the reflected pulse activates the other input. If the inputs are in the correct state, the XOR gate triggers the release of poison, potentially killing Tiger. For further details, see the main text.}
  \label{Setup}
\end{figure}

If time dilation applied only to clocks based on light propagation, a mechanical clock and a light clock that are perfectly synchronized in the inertial frame in which both are at rest would no longer be synchronized in a frame where both are in motion at the same constant velocity, because time-dilation would not apply to the mechanical clock. In order to refute this claim, we introduce the imaginary ``Sync-or-Die clock" shown in Fig.~\ref{Setup} inspired by Schrödinger's cat thought experiment\cite{Schrodinger}. A mechanical stopwatch has a contact that sends a signal (wire 1) to a laser when it is started, which results in a light pulse being emitted upward in the direction of a mirror located at a distance $h=4,496,886,870\:{\rm m}$. After the pulse has been emitted, an actuator automatically shifts the laser to the left and a light detector takes its place. This happens before the pulse comes back. The stopwatch has another contact that sends a signal (wire 2) to one input of an XOR gate after $30\:{\rm s}$ have elapsed. After the pulse reflects onto the mirror, it comes back to the detector that sends a signal (wire 3) to the other input of the gate. The output of the gate is connected (wire 4) to a device that can release poison and kill Tiger if triggered. The operation of the Sync-or-Die clock is clearly illustrated through detailed animations\cite{Rousseau}.

\begin{figure}[h]
  \centerline{\includegraphics[width=0.4\textwidth]{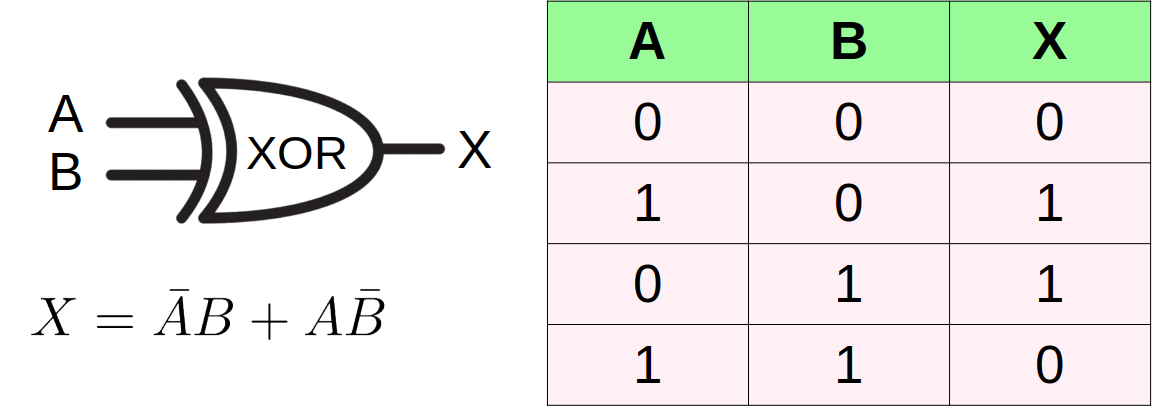}}
  \caption{The XOR gate’s truth table. The output is high if one input is high and the other is low, otherwise the output is low. This table defines the logic that determines whether the poison release mechanism is triggered in the Sync-or-Die clock.}
  \label{XOR-TruthTable}
\end{figure}

The XOR gate's truth table is given in Fig.~\ref{XOR-TruthTable}. The state of the output $X$ depends on the states of both inputs $A$ and $B$, and is computed as $X=\bar AB+A\bar B$. This results in the state of the output being high if one of the inputs is high and the other is low, otherwise the output is low.

\begin{figure}[h]
  \centerline{\includegraphics[width=0.5\textwidth]{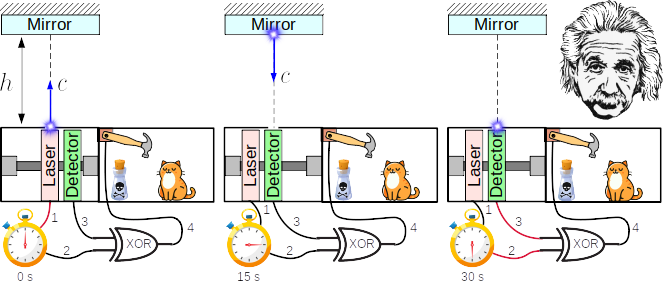}}
  \caption{Einstein observes the Sync-or-Die clock from the inertial frame where the device is at rest and concludes that Tiger survives the experiment. For further details, see the main text and Ref.~\cite{Rousseau} for detailed anmations.}
  \label{ThoughtExperiment-1}
\end{figure}

We now consider the first part of a thought experiment where the Sync-or-Die clock and Einstein are at rest in an inertial frame (Fig.~\ref{ThoughtExperiment-1}). As he starts the stopwatch, he observes the laser pulse being emitted vertically upward at the speed of light, $c=299,792,458\:{\rm m/s}$, and the detector taking the place of the laser. The pulse reaches the mirror after a time $t_\uparrow=\frac{h}c=15\:{\rm s}$ and is reflected downward. 15 s later, the pulse reaches the detector which sends a signal to one of the gate's inputs. At the same time, the stopwatch indicates 30 s and sends a signal to the gate's other input. Since both gate's inputs are in identical states at all times, no signal is ever present on the output and Einstein concludes that Tiger survives the experiment.

\begin{figure}[h]
  \centerline{\includegraphics[width=0.5\textwidth]{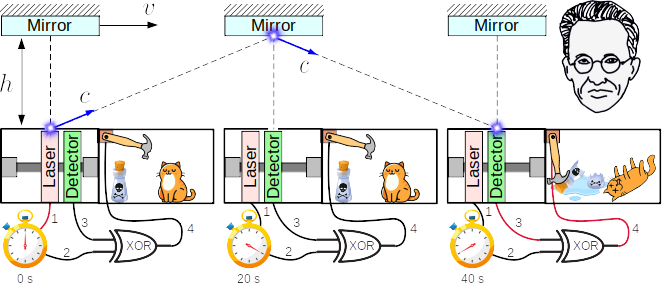}}
  \caption{Schrödinger observes the Sync-or-Die clock from an inertial frame where the device moves at constant velocity and concludes that Einstein's cat does not survive the experiment. For further details, see the main text and Ref.~\cite{Rousseau} for detailed animations.}
  \label{ThoughtExperiment-2}
\end{figure}

The second part of the thought experiment is where Schrödinger accepts the fact that the speed of light is the same in all inertial frames, as postulated by Einstein\cite{Einstein} and experimentally supported, but he considers the hypothesis that time dilation applies to light only, not to massive objects. He is in an inertial frame and he observes Einstein and his Sync-or-Die clock moving at a velocity $v=\frac{\sqrt{7}c}{4}$ (Fig.~\ref{ThoughtExperiment-2}). As Einstein starts the stopwatch, Schrödinger observes the pulse being emitted along an oblique path at the speed of light, $c=299,792,458\:{\rm m/s}$, after what the detector takes the place of the laser. It takes a time $t^\prime_\uparrow$ for the pulse to travel from the laser to the mirror over an oblique distance $ct^\prime_\uparrow$, while the Sync-or-Die clock moves a horizontal distance $vt^\prime_\uparrow$. Applying the Pythagorean theorem, the time $t_\uparrow^\prime$ needed for the pulse to travel from the laser to the mirror must satisfy the equation:
\begin{equation}
    \nonumber (ct^\prime_\uparrow)^2=(vt^\prime_\uparrow)^2+h^2\quad\Rightarrow\quad t_\uparrow^\prime=\frac{h}{\sqrt{c^2-v^2}}=20\:{\rm s}
\end{equation}
This is the time that Schrödinger observes on his own watch. According to the hypothesis, time dilation does not apply to the stopwatch, so he concludes that when the pulse arrives to the mirror the stopwatch should indicate the same time as the time he observed, namely 20~s. By symmetry, it takes the same amount of time for the pulse to travel from the mirror to the detector, so this event should happen when the stopwatch indicates 40 s. At that moment, the detector sends a signal to the gate. However, Schrödinger knows that the stopwatch was set to send a signal to the gate earlier, when it was indicating 30~s. Since the two gate's inputs received signals at different times, Schrödinger concludes that the gate's output triggered the release of poison and that Tiger is dead.

It is clear that Einstein and Schrödinger's conclusions contradict each other, leading to a paradox.

\section{Why Tiger Must Survive: Resolving the Paradox}
Fortunately, Schrödinger is a skilled physicist who understands that Tiger’s fate cannot depend on the choice of reference frame. From his point of view, the light pulse travels along a longer diagonal path and returns to the photodetector after 40 seconds. Meanwhile, the stopwatch is programmed to send the killing signal after 30 seconds of proper time. If Schrödinger assumes that the stopwatch runs “normally,” unaffected by motion, it should send the signal well before the light pulse returns, and therefore Tiger should die.

But this conclusion creates a contradiction: in Einstein’s frame, the pulse inhibits the stopwatch signal exactly at 30 seconds, and the cat survives. Both observers must agree on the final outcome, because causality is preserved in all inertial frames. Schrödinger realizes that the resolution must lie in re-examining his hypothesis. If, as opposed to his initial hypothesis, the stopwatch experiences time dilation just like the light clock does, then although it is set to trigger at 30 seconds in its own frame, this interval corresponds to 40 seconds in Schrödinger’s frame. In other words, Schrödinger sees the stopwatch running slow. This means that the killing signal is not sent at the 30-second mark in his frame, but only after 40 seconds, exactly when the light pulse arrives to inhibit it.

The apparent paradox is thus resolved by recognizing that time dilation applies universally, not just to clocks based on light. Any clock, mechanical, biological, or radioactive, will exhibit time dilation when in motion. Once Schrödinger accepts that the stopwatch ticks more slowly from his perspective, he sees that it sends the signal just in time to be inhibited by the returning light pulse. Tiger survives in both frames, and the consistency of Special Relativity is preserved.

\section{Conclusion}
In this paper, we introduced Einstein's cat thought experiment involving the Sync-or-Die clock to address the misconception that time dilation only applies to light clocks. By analyzing the fate of Tiger in both the rest frame and a moving frame, we showed that the apparent contradiction is resolved by recognizing the universality of time dilation. This not only saves Tiger but also provides a pedagogical tool for clarifying the fundamental principle of Special Relativity. In the happy ending, Tiger survives, proving that time dilation is indeed universal and independent of the clock’s mechanism.

Thought experiments have long played a central role in both the development and teaching of physics, particularly in special relativity. Classic examples such as the twin paradox and Einstein’s train thought experiment have been used to challenge intuition and highlight key conceptual features of relativistic spacetime. As Lattery emphasizes\cite{Lattery}, thought experiments in education serve not only to illustrate abstract ideas but also to confront and clarify misconceptions. The present scenario, ``Einstein’s Cat", contributes to this tradition by directly addressing the common misconception that time dilation is an artifact of light clocks alone. By linking Tiger’s survival to the universality of time dilation across different types of clocks, the thought experiment invites deeper reflection on the foundations of relativity and reinforces the conceptual robustness of the theory.

\section{Acknowledgments}
I would like to thank Jean-Marc Lévy-Leblond for useful discussions. I would also like to acknowledge the inspiration for my work, which is encapsulated in my belief: ``The hardest thing to understand in the Theory of Relativity is why so many people want to believe it is wrong". This sentiment motivates my ongoing efforts to address and clarify misconceptions in the field of relativity.

\end{document}